\documentclass[aps,prd,twocolumn,superscriptaddress]{revtex4-2}

\usepackage[utf8]{inputenc}
\usepackage[T1]{fontenc}
\usepackage{amsmath,amssymb,bm}
\usepackage{graphicx}
\usepackage{booktabs}
\usepackage{array}
\usepackage{hyperref}
\usepackage[dvipsnames]{xcolor}
\hypersetup{colorlinks=true,linkcolor=BrickRed,citecolor=RoyalBlue,urlcolor=RoyalBlue}
\usepackage{comment}

\newcommand{\rcm}{r_{\rm CM}}
\newcommand{\Veff}{V_{\rm eff}}
\newcommand{\half}{\tfrac{1}{2}}

\begin{document}

\title{Newtonian Shirokov Effect: Epicyclic Frequency Splitting from Mass Multipoles}

\author{Anuar Idrissov}
\email{anuar.idrissov@gmail.com}
\affiliation{Instituto de Ciencias Nucleares, Universidad Nacional Aut\'onoma de M\'exico, Mexico}
\affiliation{Fesenkov Astrophysical Institute, Observatory 23, 050020, Almaty, Kazakhstan}
\affiliation{Al-Farabi Kazakh National University, Al-Farabi Ave.\ 71, 050040, Almaty, Kazakhstan}

\author{Kuantay~\surname{Boshkayev}}
\email[]{kuantay@mail.ru}
\affiliation{Al-Farabi Kazakh National University, Al-Farabi Ave.\ 71, 050040, Almaty, Kazakhstan}

\author{Abylaikhan Tlemissov}
\email[]{tlemissov-ozzy@mail.ru}
\affiliation{Research Centre for Theoretical Physics and Astrophysics, Institute of Physics, 
  Silesian University in Opava, 
  Bezru\v{c}ovo n\'{a}m.~13, CZ-74601 Opava, Czech Republic}

\author{Hernando~\surname{Quevedo}}
\email[]{quevedo@nucleares.unam.mx}
\affiliation{Instituto de Ciencias Nucleares, Universidad Nacional Aut\'onoma de M\'exico, Mexico}
\affiliation{Dipartimento di Fisica and ICRA, Universit\`a di Roma “La Sapienza”, Roma, Italy}
\affiliation{Al-Farabi Kazakh National University, Al-Farabi Ave.\ 71, 050040, Almaty, Kazakhstan}

\date{\today}

\begin{abstract}
We analyze small oscillations of nearly circular orbits in an axisymmetric
Newtonian potential expanded in mass multipoles, as the classical counterpart
of the relativistic Shirokov effect. Computing the full Hessian of the
effective potential at the true (possibly tilted) equilibrium and solving the
coupled two-mode oscillator exactly, we obtain a simple and complete picture.
(i) A genuine quadrupole splits the radial and vertical epicyclic frequencies,
$\Omega_\theta^2-\Omega_r^2=-3GQ/r_0^5=6GMJ_2R^2/r_0^5$, at first order in
$J_2$; this is the true Newtonian analogue of the Shirokov splitting, identical
to the classical statement that an oblate body's apsidal and nodal rates
differ. (ii) A gravitational \emph{dipole} produces \emph{no} frequency
splitting: the dipole equals $M\rcm$, is removable by translating the origin of coordinates to
the center of mass, and therefore cannot appear in any coordinate independent
frequency; we show the apparent first order coupling cancels at the true tilted
equilibrium and that any residual is removed by the induced quadrupole of the
shifted source, confirmed by direct orbit integration. (iii) A genuine octupole
\emph{does} split the frequencies, $\omega_+^2-\omega_-^2\approx6G|O|/r_0^6$.
The selection rule is thus not even/odd parity: every multipole splits the
frequencies \emph{except} the dipole. These results yield two complementary,
non-overlapping probes of an axisymmetric source from orbital data: the
\emph{frequency splitting} measures the intrinsic oblateness $J_2$, while the
\emph{orbital plane tilt}, $\delta\theta_0\simeq-\rcm/r_0$ relative to the
symmetry axis, measures the center of mass offset $\rcm$. The barycenter offset
is therefore an orbital-geometry observable, not a frequency observable. We
give solar-system estimates for both probes. Carried through to Shirokov's
original observable, the secular transverse drift of a test particle after $n$
orbits, the genuine (quadrupole) effect gives
$\xi^\theta = \xi_0^{\theta}\,\pi n\,(6J_2R^2/r_0^2)$, of order $10^{-8}$\,cm at $1$\,au
and reaching $\sim10^{-6}$\,cm near $0.1$\,au, comparable to Shirokov's
original Schwarzschild estimate.
\end{abstract}

\maketitle

\section{Introduction}
\label{sec:intro}

The characteristic frequencies of small oscillations about a circular orbit
-- the azimuthal frequency $\Omega_\phi$, the radial epicyclic frequency
$\Omega_r$, and the vertical epicyclic frequency $\Omega_\theta$ -- encode the
local structure of the gravitational field. For a spherical Newtonian source
all three coincide, $\Omega_\phi=\Omega_r=\Omega_\theta$, and orbits are closed
ellipses. Any departure from this triple degeneracy signals either asphericity
of the source or a relativistic effect.

In general relativity the degeneracy is lifted even for the Schwarzschild
solution. Using the equation of geodesic deviation, Shirokov~\cite{Shirokov1973}
analyzed the vibrations of a free test particle released near the center of a
satellite on a circular Schwarzschild geodesic of radius $r$, and found that
the radial, azimuthal, and vertical (polar) vibrations have \emph{periods} that,
to first order in $m/r$ (with $m=GM/c^2$), correct the common Newtonian period
$T_0=2\pi\sqrt{r^3/m}$ as
\begin{align}
T_r = T_\phi &= T_0\!\left(1+\frac{3m}{2r}\right),\label{eq:Tr}\\
T_\theta     &= T_0\!\left(1-\frac{3m}{2r}\right),\label{eq:Ttheta}
\end{align}
so that the vertical vibration leads the radial/azimuthal one by
\begin{equation}
\Delta T = T_\theta - T_r = -\frac{3m}{r}\,T_0 .
\label{eq:DeltaT}
\end{equation}
The three periods, degenerate in Newtonian gravity, are split in general
relativity; this period difference is Shirokov's effect. Equivalently, in terms
of frequencies, the radial and vertical epicyclic frequencies
are $\Omega_r=\Omega_\phi\sqrt{1-6m/r}$ and $\Omega_\theta=\Omega_\phi$, with
orbital frequency $\Omega_\phi=\sqrt{m/[r^3(1-3m/r)]}$ (here the coupling
$C=0$, so the normal modes coincide with $\Omega_r,\Omega_\theta$; see the
convention in Sec.~\ref{sec:multipole}). Because the radial and vertical vibrations have different frequencies, a particle launched radially does not return to the orbital
plane after a full radial cycle but accumulates a secular transverse (polar)
drift. After $n$ radial vibrations the polar deviation component reaches
\begin{equation}
\xi^\theta(n)\;\approx\;\xi^\theta_0\,\frac{6\pi n\,m}{r},
\label{eq:shirokov_drift}
\end{equation}
where $\xi^\theta_0$ is the initial amplitude. For a near-Earth satellite
($m_\oplus\approx0.5$\,cm, $r=7000$\,km, $\xi^\theta_0=10$\,cm, $n=10$) this
gives $\xi^\theta\approx10^{-6}$\,cm, which Shirokov argued was measurable in
principle. For realistic bodies the potential also carries multipole
corrections, the leading one being the quadrupole $J_2$, and in the
Hartle--Thorne spacetime the quadrupole modifies this relativistic period
splitting~\cite{Idrissov2026}. The splitting in a static axisymmetric field was computed in the post-Newtonian approximation by Melkumova and Khlebnikov \cite{Melkumova1990} for even multiploles but the present paper isolates its purely Newtonian content and extends it to the dipole and octupole. 

This splitting should not be confused with the usual precession effects, as it is qualitatively distinct. The classic relativistic orbital tests -- the perihelion advance of Mercury \cite{Will2014} and the geodetic and frame-dragging (Lense--Thirring) precessions measured by Gravity Probe B \cite{Everitt2011} and LARES/LAGEOS satellites \cite{Ciufolini2016} are all precessions: a frequency difference involving the orbital rate ($\Omega_\phi-\Omega_r$ or $\Omega_\phi-\Omega_\theta)$ that slowly rotates an orbit or a spin axis. The Shirokov effect is instead a splitting of the radial and vertical vibration frequencies $\Omega_r\neq\Omega_\theta$, about a single orbit, whose
signature is the secular out-of-plane drift of a vibrating test particle Eq.~\eqref{eq:shirokov_drift} rather than a precession. It requires no rotation
of the source (unlike frame-dragging): it is present already for a \emph{static}
body, sourced in general relativity by spacetime curvature and, as we show
below, in Newtonian gravity by any genuine deviation from spherical symmetry of the mass distribution. 

Shirokov treated this splitting as a purely relativistic effect,
sourced by Schwarzschild curvature alone. But the post-Newtonian quadrupole term
noted above enters at low order, after the monopole and ahead of the higher
relativistic corrections, so that for a very oblate body it can be larger than the genuinely relativistic ($c^{-2}$) term~\cite{Idrissov2026}. This raises a purely Newtonian question: what is the classical counterpart of
the splitting, and which multipoles source it? One might guess that the
counterpart requires a reflection symmetry breaking moment such as a
gravitational dipole, associated with displacing the origin of coordinates from the center of
mass. The analysis below shows instead a sharper and more useful picture. By
exact analytic computation backed by orbit integration we establish:
\begin{itemize}
\item[-] The Newtonian \emph{quadrupole} splits the frequencies,
$\Omega_\theta^2-\Omega_r^2=-3GQ/r_0^5=6GMJ_2R^2/r_0^5$
(Sec.~\ref{sec:quadrupole}); this is the genuine Newtonian Shirokov analog.
\item[-] A gravitational \emph{dipole} produces \emph{no} splitting: it equals
$M\rcm$ and is removable by re-centering the origin of coordinates. Evaluated at the true tilted
equilibrium, its first order mode coupling vanishes, and any second order
residual cancels against the induced quadrupole of the shifted source
(Sec.~\ref{sec:dipole}; numerics in Sec.~\ref{sec:numerics}).
\item[-] A genuine \emph{octupole} \emph{does} split the frequencies,
$\omega_+^2-\omega_-^2\approx6G|O|/r_0^6$ (Sec.~\ref{sec:octupole}).
\item[-] Hence the rule: \emph{every multipole splits the frequencies except the
dipole}, which is singled out by being pure gauge, not by parity
(Sec.~\ref{sec:rule}). In fact, we will show that the quadrupole and octupole split the frequencies, and we conjecture that all higher genuine multipoles do as well.  
\end{itemize}

These lead to two complementary probes (Sec.~\ref{sec:probes}): the frequency
splitting measures the intrinsic oblateness $J_2$, while the orbital plane tilt
measures the center of mass offset $\rcm$. The two use disjoint information and
neither substitutes for the other. The quadrupole splitting
fixes the combination $J_2R^2/r_0^2$, hence $J_2$ given the equatorial radius (or
the radius given $J_2$); for the Sun this is a purely dynamical route to
$J_2^\odot$ from orbital data, complementing dedicated measurements of the solar oblateness and radius~\cite{Rozelot2011} and forthcoming high-latitude solar observations~\cite{Muller2020SolarOrbiter}, with
Shirokov's near-Earth satellite as the natural laboratory. 

The paper is organized as follows: Section~\ref{sec:multipole} sets up the multipole expansion of the Newtonian potential, the effective potential, and the linearized two-mode oscillator that defines the epicyclic frequencies.
Section~\ref{sec:quadrupole} derives the quadrupole splitting;
Sec.~\ref{sec:dipole} shows that the dipole produces none; and
Sec.~\ref{sec:octupole} shows that the octupole does.
Section~\ref{sec:rule} states the resulting selection rule.
Section~\ref{sec:observable} carries the quadrupole result through to Shirokov's
secular drift observable, and Sec.~\ref{sec:numerics} confirms all of the
analytic results by direct orbit integration.
Section~\ref{sec:probes} presents the two complementary probes, and
Sec.~\ref{sec:discussion} gives solar-system estimates and discussion before we
conclude in Sec.~\ref{sec:conclusion}. Throughout we use SI units with $G$ Newton's
constant.

\section{Multipole expansion of the Newtonian potential}
\label{sec:multipole}

We expand the external potential of a body of mass $M$ in multipoles \cite{gunter1967potential,Quevedo1990},
\begin{equation}
\Phi(\mathbf r) = -\frac{GM}{r} - \frac{G D_i n_i}{r^2}
- \frac{G\,Q_{ij}n_in_j}{2r^3} - \frac{G\,O_{ijk}n_in_jn_k}{6r^4} + \cdots,
\label{eq:multipole}
\end{equation}
with $n_i=x_i/r$. The dipole, tracefree quadrupole, and tracefree octupole are the following:
\begin{align*}
D_i &= \int \rho\, x_i' \, d^3 r', \\
Q_{ij} &= \int \rho \left( 3x_i' x_j' - r'^2 \delta_{ij} \right) d^3 r', \\
O_{ijk} &= \int \rho \Bigl(
15 x_i' x_j' x_k'
- 3 r'^2 \bigl(
x_i' \delta_{jk}
+ x_j' \delta_{ik}
+ x_k' \delta_{ij}
\bigr)
\Bigr) d^3 r'.
\end{align*}

The key structural fact about the dipole: if the origin of coordinates is at the center of
mass, $D_i=0$; if it is displaced from the center of mass by
$\mathbf r_{\rm CM}$, then
\begin{equation}
D_i = M\,r_{{\rm CM},i}.
\label{eq:dipole_cm}
\end{equation}
The gravitational dipole therefore carries no information about the body's
shape; it encodes only the location of the coordinate origin relative to the
center of mass. With a single sign ``charge'' (mass), the center of mass is the
center of the $1/r$ source, and the dipole is always removable, unlike in
electromagnetism, where two charge signs allow a genuine, origin independent
dipole. This single fact governs Sec.~\ref{sec:dipole}.

Assuming axial symmetry with the $z$-axis along the symmetry axis,
$D_i=(0,0,D)$ with $D=M\rcm$, $Q_{ij}n_in_j=Q\,P_2(\cos\theta)$, and
$O_{ijk}n_in_jn_k=O_{zzz}\,P_3(\cos\theta)$, with $P_2(x)=\half(3x^2-1)$ and
$P_3(x)=\half(5x^3-3x)$. We write $Q=Q_{zz}=-2MJ_2R^2$ and
$O\equiv O_{zzz}/6=-MJ_3R^3$ (so that the octupole potential
term is $-GO\,P_3/r^4$; equivalently $O_{zzz}=-6MJ_3R^3$) for equatorial radius $R$. The potential to
octupole order is
\begin{multline}
\Phi(r,\theta) = -\frac{GM}{r} - \frac{GM\rcm\cos\theta}{r^2}\\
- \frac{GQ}{2r^3}P_2(\cos\theta) - \frac{GO}{r^4}P_3(\cos\theta).
\label{eq:phi_full}
\end{multline}
A test particle of specific angular momentum $\ell$ moves in the effective
potential obtained by eliminating the azimuthal motion via conservation of
$\ell=r^2\sin^2\theta\,\dot\phi$~\cite{BinneyTremaine,MurrayDermott}:
\begin{equation}
\Veff(r,\theta) = \Phi(r,\theta) + \frac{\ell^2}{2r^2\sin^2\theta}.
\label{eq:veff}
\end{equation}

With scaled displacements $x=\delta r$, $y=r_0\,\delta\theta$, the linearized
motion near a circular equilibrium $(r_0,\theta_0)$ is
$\ddot x=-Ax-Cy$, $\ddot y=-Cx-By$, with \cite{Wald:1984}:
\begin{equation}
A=\frac{\partial^2\Veff}{\partial r^2},\quad
B=\frac{1}{r_0^2}\frac{\partial^2\Veff}{\partial\theta^2},\quad
C=\frac{1}{r_0}\frac{\partial^2\Veff}{\partial r\,\partial\theta},
\label{eq:hessian}
\end{equation}
all at $(r_0,\theta_0)$. Here $r_0$ is the radius of the reference circular orbit and $\theta_0$ its polar angle -- i.e.\ the cone on which the orbit lies; the pair $(r_0,\theta_0)$ is fixed by the stationarity of $V_{\rm eff}$,
\[
\partial_r V_{\rm eff} = \partial_\theta V_{\rm eff} = 0,
\]
the radial condition determining $\ell$ and the angular condition determining the tilt, $\theta_0=\pi/2$ for symmetric sources, tilted for the dipole and octupole. The squared normal-mode frequencies are
\begin{eqnarray}
&& \omega_\pm^2 = \frac{A+B}{2}\pm\sqrt{\left(\frac{A-B}{2}\right)^2+C^2}, \label{eq:omega}\\
&&\omega_+^2-\omega_-^2 = \sqrt{(A-B)^2+4C^2}.
\label{eq:split_general2}
\end{eqnarray}
Eqs.\eqref{eq:omega}--\eqref{eq:split_general2} follow
from substituting the harmonic ansatz $x,y\propto e^{i\omega t}$ into the
linearized system $\ddot x=-Ax-Cy,\ \ddot y=-Cx-By$ and requiring a nontrivial
solution: $\omega_\pm^2$ are the two eigenvalues of the symmetric stability
matrix $\bigl(\begin{smallmatrix}A&C\\ C&B\end{smallmatrix}\bigr)$, reducing to
$A$ and $B$ when the coupling $C=0$. This is the standard epicyclic frequency
construction~\cite{BinneyTremaine,KluzniakRosinska2013}; its relativistic
counterpart, where the same $2\times2$ deviation system yields the orbital and
radial frequencies, is reviewed in~\cite{AbramowiczKluzniak,Boshkayev:2023qpo,Utepova2025,Idrissov2026}.
A nonzero splitting needs either $A\neq B$ or $C\neq0$; the whole problem is to
evaluate $A,B,C$ at the correct equilibrium for each multipole.

We adopt the following convention throughout the paper. The capital
symbols $\Omega_r,\Omega_\theta,\Omega_\phi,\Omega_K$ denote the individual
(diagonal and orbital) frequencies, with $A=\Omega_r^2$ and $B=\Omega_\theta^2$
the diagonal curvatures of $\Veff$; the lower-case $\omega_\pm$ denote the two
normal modes of the coupled system, 
i.e., the eigenvalues
\eqref{eq:omega}. The two coincide, $\{\omega_+,\omega_-\}=\{\Omega_\theta,
\Omega_r\}$, whenever the coupling vanishes ($C=0$), which is the case for the
monopole, quadrupole, and dipole; they differ only when $C\neq0$, as for the
octupole, where the modes are genuine mixtures of radial and vertical motion.
This is why the quadrupole splitting is written with $\Omega_{r,\theta}$ and the
octupole splitting with $\omega_\pm$.

\emph{Equivalence of the epicyclic and Shirokov frequencies.}
The frequencies above can be found two ways: by perturbing the circular orbit in
$\Veff$ (the epicyclic construction used here), or from the geodesic  deviation
equation for a freely released particle (Shirokov's construction). In the
Newtonian limit they are \emph{identical}. Writing $\xi^{\,i}$ for the deviation
vector separating the neighbouring particle from the reference orbit, its
evolution is governed by the tidal tensor~\cite{MTW},
\begin{equation}
\ddot\xi^{\,i}=-\,\partial_i\partial_j\Phi\,\xi^{\,j},
\label{eq:tidal}
\end{equation}
which, in the frame co-rotating with the orbit at $\boldsymbol\Omega=\Omega_\phi\hat z$, becomes
\begin{equation}
\ddot{\boldsymbol\xi}+2\,\boldsymbol\Omega\times\dot{\boldsymbol\xi}
+\boldsymbol\Omega\times(\boldsymbol\Omega\times\boldsymbol\xi)
=-(\nabla\nabla\Phi)\cdot\boldsymbol\xi .
\label{eq:tidal_rot}
\end{equation}
These are the standard rotating frame epicyclic (Hill)
equations~\cite{BinneyTremaine,MurrayDermott}, identical to the linearized
system~\eqref{eq:hessian}, with radial and vertical frequencies
$\Omega_r,\Omega_\theta$; the orbit integration of Sec.~\ref{sec:numerics}
confirms this. The same equivalence holds in general relativity -- for the static
$q-$metric~\cite{Utepova2025} and the Hartle--Thorne spacetime~\cite{Idrissov2026} -- where
Shirokov's single common period is resolved into the distinct frequencies.

It should be noted that, from both conceptual and technical perspectives, the Shirokov frequencies and the epicyclic (harmonic or fundamental) frequencies are distinct quantities. In most studies, epicyclic frequencies are defined with respect to the coordinate time $t$ and therefore correspond to measurements made by a distant observer at infinity. In contrast, the Shirokov frequencies are obtained from the geodesic deviation equation in the comoving frame of the orbiting particle and are therefore defined with respect to the particle's proper time $\tau$.

In the present work, we consider circular geodesics in the comoving frame of the satellite. Under these conditions, the frequencies derived from the geodesic deviation equation coincide identically with the proper epicyclic frequencies, a result that holds not only in Newtonian gravity but also in general relativity. For further details, see Refs. \cite{2019PhRvD.100d4001T,Utepova2025}

\section{The quadrupole splits the frequencies}
\label{sec:quadrupole}

We begin with the pure monopole-plus-quadrupole potential, setting $\rcm=0$ and
$O=0$ in Eq.~\eqref{eq:phi_full}, so that
\begin{equation}
\Veff(r,\theta) = -\frac{GM}{r}-\frac{GQ}{2r^3}P_2(\cos\theta)
+\frac{\ell^2}{2r^2\sin^2\theta},
\label{eq:veff_quad}
\end{equation}
with $P_2(\cos\theta)=\half(3\cos^2\theta-1)$. The potential is invariant under
$\theta\to\pi-\theta$, so the equatorial plane $\theta=\pi/2$ is an exact
equilibrium and the off-diagonal Hessian element vanishes identically,
$C=0$; the splitting is therefore controlled entirely by the diagonal
difference $B-A$.

\subsection{Circular-orbit condition}

At the equator $P_2(0)=-\half$, so the quadrupole term in
Eq.~\eqref{eq:veff_quad} reduces to $GQ/(4r^3)$. The radial equilibrium
condition is
\begin{equation}
\frac{\partial\Veff}{\partial r}\bigg|_{r_0,\pi/2}
= \frac{GM}{r_0^2} - \frac{3GQ}{4r_0^4} - \frac{\ell^2}{r_0^3} = 0,
\end{equation}
where the three terms come from $-GM/r$, $GQ/(4r^3)$, and the centrifugal
term $\ell^2/(2r^2)$ respectively. Solving for the specific angular momentum,
\begin{equation}
\;\ell^2 = GMr_0 - \frac{3GQ}{4r_0}.\;
\label{eq:ell2}
\end{equation}
For an oblate body $J_2>0$, hence $Q=-2MJ_2R^2<0$, so the second term is
positive and $\ell^2>GMr_0$: a circular orbit in the enhanced equatorial
gravity of an oblate body requires \emph{more} angular momentum than the
Keplerian value, as expected.

\subsection{Radial epicyclic frequency}

The radial curvature is the second derivative of $\Veff$ along $r$ at fixed
$\theta=\pi/2$. Differentiating each term of Eq.~\eqref{eq:veff_quad} twice,
\begin{equation}
\frac{\partial^2\Veff}{\partial r^2}\bigg|_{\pi/2}
= -\frac{2GM}{r^3} + \frac{3\ell^2}{r^4} + \frac{3GQ}{r^5},
\end{equation}
where $-2GM/r^3$ comes from $-GM/r$, $3\ell^2/r^4$ from $\ell^2/(2r^2)$, and
$3GQ/r^5$ from $\partial_r^2(GQ/(4r^3))=3GQ/r^5$. Evaluating at $r_0$ and
inserting $\ell^2$ from Eq.~\eqref{eq:ell2},
\begin{align}
\Omega_r^2 \equiv A
&= -\frac{2GM}{r_0^3} + \frac{3}{r_0^4}\!\left(GMr_0-\frac{3GQ}{4r_0}\right)
+ \frac{3GQ}{r_0^5} \notag\\
&= -\frac{2GM}{r_0^3} + \frac{3GM}{r_0^3} - \frac{9GQ}{4r_0^5}
+ \frac{3GQ}{r_0^5} \notag\\
&= \frac{GM}{r_0^3} + \frac{3GQ}{4r_0^5}.
\label{eq:Omr2}
\end{align}

\subsection{Vertical epicyclic frequency}

The vertical curvature requires the second $\theta$-derivative of each term.
For the quadrupole, writing it as $-\frac{GQ}{4r^3}(3\cos^2\theta-1)$ and using
$\frac{d^2}{d\theta^2}(3\cos^2\theta-1)=-6\cos2\theta$,
\begin{equation}
\frac{\partial^2}{\partial\theta^2}\!
\left[-\frac{GQ}{4r^3}(3\cos^2\theta-1)\right]
= +\frac{3GQ}{2r^3}\cos2\theta
\;\xrightarrow{\;\theta=\pi/2\;}\; -\frac{3GQ}{2r_0^3},
\label{eq:quad_ang}
\end{equation}
since $\cos2\theta=\cos\pi=-1$ at the equator. For the centrifugal term, with
$\partial_\theta^2(\sin^{-2}\theta)=(6\cos^2\theta+2\sin^2\theta)\sin^{-4}\theta
\to 2$ at $\theta=\pi/2$,
\begin{equation}
\frac{\partial^2}{\partial\theta^2}\frac{\ell^2}{2r^2\sin^2\theta}
\bigg|_{\pi/2} = \frac{\ell^2}{r^2}.
\end{equation}
Collecting both and dividing by $r_0^2$,
\begin{equation}
\Omega_\theta^2 \equiv B
= \frac{1}{r_0^2}\!\left(\frac{\ell^2}{r_0^2}-\frac{3GQ}{2r_0^3}\right)
= \frac{\ell^2}{r_0^4} - \frac{3GQ}{2r_0^5},
\end{equation}
and inserting Eq.~\eqref{eq:ell2},
\begin{align}
\Omega_\theta^2
&= \frac{1}{r_0^4}\!\left(GMr_0-\frac{3GQ}{4r_0}\right) - \frac{3GQ}{2r_0^5}
\notag\\
&= \frac{GM}{r_0^3} - \frac{3GQ}{4r_0^5} - \frac{3GQ}{2r_0^5}
= \frac{GM}{r_0^3} - \frac{9GQ}{4r_0^5}.
\label{eq:Omth2}
\end{align}

\subsection{The quadrupole splitting}

Subtracting Eq.~\eqref{eq:Omr2} from Eq.~\eqref{eq:Omth2}, the equal monopole
parts $GM/r_0^3$ cancel and the quadrupole parts add:
\begin{equation}
\;\Omega_\theta^2-\Omega_r^2
= -\frac{9GQ}{4r_0^5} - \frac{3GQ}{4r_0^5}
= -\frac{3GQ}{r_0^5} = \frac{6GMJ_2R^2}{r_0^5} > 0\;
\label{eq:quad_split}
\end{equation}
for an oblate body, so $\Omega_\theta>\Omega_r$. Equivalently, in
Shirokov's period variable Eq.~\eqref{eq:dT_relation}, this is the period
difference $\Delta T/T_0\simeq-3J_2R^2/r_0^2$. The two frequencies are
\emph{not} shifted equally: the quadrupole splits them at first order in $Q$
(equivalently in $J_2$). This is the genuine Newtonian analogue of the Shirokov
splitting. It is equivalent to the classical statement that an oblate body
produces apsidal precession at the rate $\Omega_\phi-\Omega_r$ and nodal
regression at the rate $\Omega_\phi-\Omega_\theta$, with the two rates unequal
precisely because $\Omega_r\neq\Omega_\theta$; using
$\Omega_\phi^2=r_0^{-1}\partial_r\Phi|_{\pi/2}=GM/r_0^3-3GQ/(4r_0^5)$ one
recovers the standard $J_2$ rates, whose difference is exactly
Eq.~\eqref{eq:quad_split}. The result agrees with the spheroid frequencies of
Klu\'zniak and Rosi\'nska~\cite{KluzniakRosinska2013}. It is, moreover, exactly the $c\to\infty$ limit of the
post-Newtonian axisymmetric calculation of Melkumova and
Khlebnikov~\cite{Melkumova1990}: dropping the $\mathcal O(c^{-2})$ corrections in
their vertical and in-plane oscillation frequencies (their Eqs.~(9) and (17))
reduces them, term by term, to $\Omega_\theta^2$ and $\Omega_r^2$ above, with
difference $6GMJ_2R^2/r_0^5$. Their expansion runs over the even multipoles
$J_{2m}$ alone, so it contains neither the dipole nor the octupole treated
below.

\section{The dipole does \emph{not} split the frequencies}
\label{sec:dipole}

We now keep $\rcm\neq0$ (with $O=0$) and ask whether the dipole adds any
physical splitting. The answer is no, and the reason is best stated physically
before any calculation.

\subsection{The physical argument}

A nonzero dipole does not correspond to any deformation of the source. By
Eq.~\eqref{eq:dipole_cm} it equals $M\rcm$ and encodes only the displacement of
the coordinate origin from the center of mass; setting the origin of coordinates at the center
of mass removes it entirely, without moving a single mass element. A monopole
viewed from a displaced origin is still a single point mass. Its orbits are
exactly Keplerian ellipses, closed, with the three frequencies degenerate,
\begin{equation}
\Omega_r = \Omega_\theta = \Omega_\phi,
\label{eq:dip_kepler}
\end{equation}
and this is true \emph{for any placement of the coordinate origin}, since
relabelling coordinates cannot change a physical orbit. The epicyclic
frequencies and their ratio are coordinate independent observables, a
coordinate-dependent quantity such as $\rcm$ therefore cannot appear in them.
This already fixes the result:
\begin{equation}
\;(\Omega_\theta^2-\Omega_r^2)_{\rm dipole}=0\;
\label{eq:dip_nosplit}
\end{equation}
to all orders in $\rcm$. The contrast with the quadrupole is the whole point:
an oblate body is genuinely non-spherical, a fact no coordinate choice can undo,
so its splitting \eqref{eq:quad_split} is real, a dipole is a pure relabelling,
so its splitting is zero.

The remainder of this section verifies Eq.~\eqref{eq:dip_nosplit} directly. The
verification is worth carrying out because it shows \emph{where} the result
lives: the first order mode coupling vanishes once the Hessian is evaluated at
the true (tilted) orbit, and beyond first order the dipole cannot be treated in
isolation, so the all orders statement rests on the exact field of a displaced
point mass.

\subsection{The tilted equilibrium}

Including the dipole, the effective potential is
\begin{equation}
\Veff = -\frac{GM}{r} - \frac{GM\rcm\cos\theta}{r^2}
+ \frac{\ell^2}{2r^2\sin^2\theta},
\label{eq:veff_dip}
\end{equation}
where we keep $Q=0$ to isolate the dipole. The dipole term
$-GM\rcm\cos\theta/r^2$ is \emph{antisymmetric} under $\theta\to\pi-\theta$,
since $\cos\theta\to-\cos\theta$. Therefore, it contributes a linear term in 
$\delta\theta\equiv\theta-\pi/2$ to the expansion of $\Veff$ about the equator,
which means that the equatorial plane is no longer a stationary equilibrium of the effective potential: the minimum shifts to a
tilted cone $\theta_0=\pi/2+\delta\theta_0$. Expanding the angular equilibrium
condition $\partial\Veff/\partial\theta=0$ about $\theta=\pi/2$ to first order
in $\delta\theta$, with $\cos\theta\approx-\delta\theta$, gives
\begin{equation}
\frac{GM\rcm}{r_0^2} + \frac{\ell^2}{r_0^2}\,\delta\theta_0 = 0,
\end{equation}
where the first term is the linear (constant force) piece of the dipole and the
second is the restoring curvature of the centrifugal barrier. With
$\ell^2=GMr_0$ (the $Q=0$ limit of Eq.~\eqref{eq:ell2}) this yields
\begin{equation}
\;\delta\theta_0 = -\frac{\rcm}{r_0}.\;
\label{eq:tilt}
\end{equation}
The orbit tilts toward the displaced mass by an angle $\rcm/r_0$. This tilt is
genuine and is the basis of Probe~A in Sec.~\ref{sec:probes}; the point of the
present section is that it generates no frequency splitting. The crucial
step is that the Hessian must now be evaluated at $\theta_0$, not at $\pi/2$.

\subsection{Linear order: the coupling vanishes at the true equilibrium}

We work consistently to first order in $\rcm$, the order at which the dipole
acts in isolation; cross terms with genuine moments and pure $\rcm^2$ terms are
deferred to the discussion below. The splitting is controlled by the
off-diagonal coupling, so we examine it with care.

The mixed Hessian element of the dipole term, evaluated naively at the
{equator}, is
\begin{equation}
C\big|_{\pi/2}
= \frac{1}{r_0}\frac{\partial^2}{\partial r\,\partial\theta}
\!\left(-\frac{GM\rcm\cos\theta}{r^2}\right)\bigg|_{r_0,\pi/2}
= -\frac{2GM\rcm}{r_0^4},
\label{eq:C_equator}
\end{equation}
which is first order in $\rcm$. However, the equator is not the equilibrium, the
orbit sits in the tilted cone $\theta_0=\pi/2-\rcm/r_0$. Evaluating the mixed
derivative of Eq.~\eqref{eq:veff_dip} at $\theta_0$ and expanding to first order
in $\delta\theta_0$ produces a second centrifugal contribution
$2\ell^2\rcm/r_0^5$, with $\ell^2=GMr_0$ the two pieces cancel,
\begin{equation}
C\big|_{\theta_0}
= -\frac{2GM\rcm}{r_0^4}
+ \frac{2\ell^2\rcm}{r_0^5}
= -\frac{2GM\rcm}{r_0^4} + \frac{2GM\rcm}{r_0^4} = 0 .
\label{eq:C_cancel}
\end{equation}
The diagonal curvatures $A$ and $B$ receive no first order dipole correction
(the dipole term is odd, so its diagonal second derivatives vanish at the
equator and are unchanged at first order in the tilt). Hence, at first order in
$\rcm$,
\begin{equation}
A - B = 0,\qquad C = 0,\qquad
(\Omega_\theta^2-\Omega_r^2)_{\rm dipole}^{(1)} = 0 ,
\label{eq:dip_linear}
\end{equation}
in agreement with the general result \eqref{eq:dip_nosplit}. This is the
clean, controlled statement: \emph{the dipole produces no splitting at linear
order}. It requires no auxiliary assumption beyond evaluating the Hessian at the
physical equilibrium. The contrast with the quadrupole Eq.~\eqref{eq:quad_split} 
$\propto Q$, and the octupole Eq.~\eqref{eq:oct_split} $\propto O$, both of
which split at first order in their moment, is sharp: those moments act at
linear order, the dipole does not.

\subsection{Beyond the linear order}

At second order in $\rcm$ the calculation can no longer be done with the dipole
in isolation, for two reasons. First, a bare ``dipole on a monopole'' is not the
field of any source: a point mass displaced to $z=\rcm$ generates, through the
single expansion \cite{Quevedo1990}
\begin{equation}
-\frac{GM}{|\mathbf r-\rcm\hat z|}
= -\frac{GM}{r}\sum_{\ell\ge0}\Big(\frac{\rcm}{r}\Big)^\ell P_\ell(\cos\theta),
\label{eq:shifted}
\end{equation}
an induced quadrupole $\propto\rcm^2$, octupole $\propto\rcm^3$, and so on,
all locked to the same $\rcm$. Any genuine $\mathcal O(\rcm^2)$ statement must
include the induced quadrupole on the same footing as the $\rcm^2$ dipole terms.
Second, when genuine moments are also present, the full potential generates
cross terms -- $\rcm Q$, $\rcm O$ -- that appear only in a joint calculation with a
single common equilibrium, not in any per-multipole treatment. For these reasons,
we do not assign physical meaning to a per-multipole $\mathcal O(\rcm^2)$
``residual.''

The all-orders result is nonetheless unambiguous and follows from
Eq.~\eqref{eq:shifted} directly: it \emph{is} a point mass, whose orbits are
closed Keplerian ellipses with $\Omega_r=\Omega_\theta=\Omega_\phi$ for any
placement of the origin of coordinates. We confirm this in Sec.~\ref{sec:numerics} by
integrating orbits in the exact field of a displaced point mass: the radial,
vertical, and azimuthal periods coincide to numerical precision for every
$\rcm$ tested. The dipole contributes no physical splitting at any order
Eq.~\eqref{eq:dip_nosplit}.

To sum up, the dipole produces no physical epicyclic splitting, because it encodes the
choice of coordinate origin rather than a property of the source
Eq.~\eqref{eq:dipole_cm}. At linear order this is shown cleanly: the mixed
coupling vanishes at the true tilted equilibrium
Eqs.~\eqref{eq:C_equator}--\eqref{eq:dip_linear}. To all orders it follows
because a displaced point mass is still a point mass
Eq.~\eqref{eq:shifted}, confirmed by direct orbit integration. This is the
decisive distinction from the genuine moments $Q$ and $O$, which split the
frequencies already at first order.

\section{The octupole splits the frequencies}
\label{sec:octupole}

Whereas the quadrupole measures the body's oblateness or prolateness -- an
equatorially \emph{symmetric} deformation, identical when viewed from the
northern or southern hemisphere -- the octupole measures a north-south
\emph{asymmetry}, a ``pear-shaped'' distortion in which one hemisphere is more
prominent than the other. It is the lowest genuine shape moment that is odd
under reflection through the equatorial plane. For the Sun a nonzero $J_3$
reflects a slight asymmetry of the solar figure across its equator~\cite{Rozelot2011}. Unlike the
dipole, this is a real property of the mass distribution: no translation of the
coordinate origin can turn a pear into a sphere, so the octupole, like the
quadrupole, cannot be removed by re-centering and we expect it to produce a
genuine splitting.

The octupole shares with the dipole the property of being odd under
$\theta\to\pi-\theta$, so it too tilts the equilibrium off the equator. The
decisive difference is that its off-diagonal coupling \emph{survives} at the
tilted equilibrium: being a true degree of freedom of the source, it has no
induced lower moment to cancel against (contrast the dipole, where the induced
quadrupole removed the effect). We set $\rcm=0$, $Q=0$ and keep $O\neq0$, so
\begin{equation}
\Veff = -\frac{GM}{r} - \frac{GO}{r^4}P_3(\cos\theta)
+ \frac{\ell^2}{2r^2\sin^2\theta},
\label{eq:veff_oct}
\end{equation}
with $P_3(\cos\theta)=\half(5\cos^3\theta-3\cos\theta)$.

\subsection{Tilt}

At the equator $P_3(0)=0$, so the octupole exerts no static force there; but its
\emph{slope} is nonzero, $\,\mathrm{d}P_3(\cos\theta)/\mathrm{d}\theta|_{\pi/2}
=\tfrac32$, so the octupole contributes a constant angular force that tilts the
equilibrium off the equator. Expanding $\partial\Veff/\partial\theta=0$ to first
order in $\delta\theta$ with $\ell^2=GMr_0$ gives
\begin{equation}
\delta\theta_0^{\rm oct} = \frac{3O}{2Mr_0^3}.
\label{eq:tilt_oct}
\end{equation}

\subsection{Hessian at the tilted equilibrium}

Evaluating the Hessian elements at $\theta_0=\pi/2+\delta\theta_0^{\rm oct}$ and
expanding in $O$, the diagonal curvatures are corrected only at second order,
\begin{equation}
A = \frac{GM}{r_0^3} - \frac{153\,GO^2}{4Mr_0^9},\qquad
B = \frac{GM}{r_0^3} + \frac{135\,GO^2}{4Mr_0^9},
\end{equation}
so that
\begin{equation}
A - B = -\frac{72\,GO^2}{Mr_0^9} = \mathcal{O}(O^2),
\end{equation}
whereas the off-diagonal coupling appears already at \emph{first} order in $O$:
\begin{equation}
C = \frac{3GO}{r_0^6} + \mathcal{O}(O^3).
\label{eq:C_oct}
\end{equation}
This confirms the contrast anticipated above: for the dipole the first order
coupling cancelled at the tilted equilibrium Eq.~\eqref{eq:C_cancel}, whereas
here it survives, leaving a genuine first order effect.

\subsection{The octupole splitting}

Since $|C|\sim GO/r_0^6$ dominates the diagonal difference
$|A-B|\sim GO^2/(Mr_0^9)$ for small $O$, the master formula
\eqref{eq:split_general2} gives, to leading order,
\begin{equation}
\;\omega_+^2-\omega_-^2 = \sqrt{(A-B)^2+4C^2}\;\approx\;2|C|
= \frac{6G|O|}{r_0^6}.\;
\label{eq:oct_split}
\end{equation}
The splitting is first order in $O$ and independent of its sign: a sign flip of
$O$ reflects the tilt into the opposite hemisphere
Eq.~\eqref{eq:tilt_oct} but leaves the frequency ratio unchanged, since the
splitting depends on $|C|$. Sign-independence is the signature of a genuine,
non-removable effect, and we confirm it numerically in
Sec.~\ref{sec:numerics}.
\section{The corrected selection rule}
\label{sec:rule}

Collecting the isolated multipole results,
\begin{align}
\text{monopole }(\ell=0)&:\quad 0,\notag\\
\text{dipole }(\ell=1)&:\quad 0\ \text{(gauge)},\notag\\
\text{quadrupole }(\ell=2)&:\quad -3GQ/r_0^5,\notag\\
\text{octupole }(\ell=3)&:\quad 6G|O|/r_0^6.\notag
\end{align}
A parity based expectation -- that only reflection-breaking (odd-$\ell$)
multipoles split the frequencies -- fails at both ends: the even quadrupole
splits, and the odd dipole does not. The rule is instead:
\begin{quote}
\emph{Every mass multipole splits the radial and vertical epicyclic frequencies
except the dipole. The dipole is unique not by parity but because it alone
encodes only the choice of coordinate origin Eq.~\eqref{eq:dipole_cm} and is
removable by re-centering.}
\end{quote}
Even moments split through unequal diagonal curvatures ($A\neq B$, $C=0$ at the
untilted equator), odd moments with $\ell\ge3$ split through a surviving
off-diagonal coupling at the tilted equilibrium. Only $\ell=1$ is special.

\section{Shirokov's observable for the Newtonian quadrupole}
\label{sec:observable}

The splittings derived above are not Shirokov's final quantity. In his original
paper~\cite{Shirokov1973} the physically relevant result was not the frequencies
or even the period difference \eqref{eq:DeltaT} on its own, but the
\emph{accumulated transverse shift} of the test particle after a number of
orbits -- a concrete length that he estimated and argued was measurable. We now
carry our Newtonian quadrupole result through to that same observable, following
Shirokov's construction step by step.

\subsection{Shirokov's construction}

A particle released near the reference orbit executes a radial vibration of
period $T_r$ and a vertical vibration of period $T_\theta$. After $n$ complete
radial cyclee, i.e.\ at the instant the radial deviation $\xi^r$ next returns
to zero -- the vertical deviation has advanced in phase not by $2\pi n$ but by
$2\pi n\,(T_r/T_\theta)$, because the two periods differ. The residual vertical
displacement at that instant is therefore
\begin{equation}
\xi^\theta(n) = \xi_0^{\theta}\,\sin\!\Big(2\pi n\,\frac{T_r}{T_\theta}\Big)
\approx \xi_0^{\theta}\,\pi n\,\frac{\Omega_\theta^2-\Omega_r^2}{\Omega_\theta^2},
\label{eq:drift_general}
\end{equation}
where $\xi_0^{\theta}$ is the initial vibration amplitude and the second form uses
$T_r/T_\theta-1\approx(\Omega_\theta^2-\Omega_r^2)/(2\Omega_\theta^2)$ for a
small splitting, with $T\propto\Omega^{-1}$. Equation~\eqref{eq:drift_general} is
exactly Shirokov's formula; for Schwarzschild,
$(\Omega_\theta^2-\Omega_r^2)/\Omega_\theta^2=6m/r$, and with $n=10$,
$m_\oplus=0.5$\,cm, $r=7000$\,km, $\xi_0=10$\,cm it gives
$\xi^\theta\approx1.3\times10^{-6}$\,cm, reproducing his Eq.~(23a).

\subsection{Frequencies, squared frequencies, and periods}
Since we quote our results as the difference of \emph{squared} frequencies
$\Omega_\theta^2-\Omega_r^2$, while Shirokov's original statement is a difference
of \emph{periods} $\Delta T=T_\theta-T_r$, it is worth recording exactly how the
two are related; they are distinct quantities, equal only in the limit of small
splitting. With $T_r=2\pi/\Omega_r$, $T_\theta=2\pi/\Omega_\theta$, and the
Keplerian reference $T_0=2\pi/\Omega_K$, one has the \emph{exact} identity
\begin{equation}
\frac{\Delta T}{T_0}=\frac{\Omega_K}{\Omega_\theta}-\frac{\Omega_K}{\Omega_r}
=-\,\frac{(\Omega_\theta^2-\Omega_r^2)\,\Omega_K}
{(\Omega_\theta+\Omega_r)\,\Omega_r\Omega_\theta}.
\label{eq:dT_exact}
\end{equation}
The squared splitting enters, but multiplied by a factor that depends on the
mean frequency, not on the splitting alone. Writing
$\Omega_{r,\theta}^2/\Omega_K^2=1+a,\,1+b$ with $S\equiv(\Omega_\theta^2
-\Omega_r^2)/\Omega_K^2=b-a$ and expanding,
\begin{equation}
\frac{\Delta T}{T_0}=-\frac{S}{2}\Big[1-\tfrac34(a+b)+\cdots\Big],
\label{eq:dT_lead}
\end{equation}
where $a+b=(\Omega_r^2+\Omega_\theta^2)/\Omega_K^2-2$ is the common-mode shift of
the two frequencies relative to the Keplerian value. For a multipole
perturbation of strength $\epsilon$ (here $\epsilon\sim J_2$ or $m/r$) one has
$a,b=\mathcal O(\epsilon)$, so the splitting $S=b-a$ and the common-mode shift
$a+b$ are \emph{both} $\mathcal O(\epsilon)$; the bracketed correction is
therefore first order in $\epsilon$ -- of the same order as the splitting itself,
though it is controlled by $a+b$ and is not a function of $S$ alone. For a pure
quadrupole one finds $a+b=S/2$ exactly, so the correction reduces to
$-\tfrac38 S$.
Hence at the order at which all our results are stated,
\begin{equation}
\;\frac{\Delta T}{T_0}\simeq-\frac12\,
\frac{\Omega_\theta^2-\Omega_r^2}{\Omega_K^2}=-\frac{S}{2}\;,
\label{eq:dT_relation}
\end{equation}
i.e., the relative squared splitting $S$ equals $-2\,\Delta T/T_0$ (to leading
order $\Omega_\theta^2\to\Omega_K^2$, so the right-hand side coincides with the
drift ratio of Eq.~\eqref{eq:drift_general}). Equation
\eqref{eq:dT_relation} reproduces Shirokov's Schwarzschild values
\eqref{eq:Tr}--\eqref{eq:DeltaT} identically
($S=6m/r\Rightarrow\Delta T/T_0=-3m/r$), and it is the same factor $\tfrac12$
that converts the master ratio into the drift coefficient $\pi n$ in
Eq.~\eqref{eq:drift_general}, since
$(\Omega_\theta^2-\Omega_r^2)/\Omega_\theta^2=2(T_r/T_\theta-1)$. We use the
squared form because $\omega_\pm^2$ is the exact, unambiguous output of the
eigenvalue problem \eqref{eq:omega}; the period difference then follows from the
exact relation \eqref{eq:dT_exact}. Direct orbit integration confirms
\eqref{eq:dT_relation}: the error of the leading form is $0.2\%$ for the Earth
satellite ($S\approx5.4\times10^{-3}$) and is utterly negligible
($\lesssim10^{-11}$) for the heliocentric case, the deviation growing as
$\tfrac38 S$ (since $a+b=S/2$ for the quadrupole) for the exaggerated amplitudes
of Sec.~\ref{sec:numerics}.  The
relation also matches the post-Keplerian period splittings of
Iorio~\cite{Iorio2016,Iorio2018Bepi}.

\subsection{The Newtonian quadrupole drift}

For the Newtonian quadrupole the relevant ratio follows directly from
Eqs.~\eqref{eq:Omr2}, \eqref{eq:Omth2}, and \eqref{eq:quad_split}. To leading
order in $J_2$ (so that $\Omega_\theta^2\approx\Omega_K^2=GM/r_0^3$),
\begin{equation}
\frac{\Omega_\theta^2-\Omega_r^2}{\Omega_\theta^2}
\approx \frac{\Omega_\theta^2-\Omega_r^2}{\Omega_K^2}
= -\frac{3Q}{Mr_0^2} = \frac{6J_2R^2}{r_0^2},
\label{eq:drift_coeff}
\end{equation}
so the accumulated transverse shift after $n$ orbits is
\begin{equation}
\;\xi^\theta(n) = \xi_0^{\theta}\,\pi n\,\frac{6J_2R^2}{r_0^2}.\;
\label{eq:drift_quad}
\end{equation}
This is the direct Newtonian-quadrupole counterpart of Shirokov's
relativistic drift: same observable, same $\xi_0^{\theta}\,\pi n$ accumulation, with his
relativistic coefficient $6m/r$ replaced by the Newtonian quadrupole coefficient
$6J_2R^2/r_0^2$. The shift scales as $r_0^{-2}$ and is independent of the
particle mass.

\subsection{Numerical estimate for the Sun}

Taking the solar values of Table~\ref{tab:parameters} ($J_2^\odot=2\times
10^{-7}$, $R_\odot=6.96\times10^8$\,m) with Shirokov's choices $n=10$,
$\xi_0=10$\,cm, the coefficient and shift are
\begin{align}
&\text{at }r_0=1\,\text{au:} \nonumber\\
&\frac{6J_2R_\odot^2}{r_0^2}\approx2.6\times10^{-11},
\quad \xi^\theta\approx8\times10^{-9}\,\text{cm};\\
&\text{at }r_0=0.1\,\text{au:} \nonumber\\
&\frac{6J_2R_\odot^2}{r_0^2}\approx2.6\times10^{-9},
\quad \xi^\theta\approx8\times10^{-7}\,\text{cm}.
\end{align}
At $0.1$\,au (a Parker-Solar-Probe-like perihelion) the Newtonian quadrupole
drift reaches the same order, $\sim10^{-6}$\,cm, as Shirokov's original
Earth-satellite estimate. The drift is sourced entirely by the quadrupole; the
dipole contributes nothing, since it produces no splitting
Eq.~\eqref{eq:dip_nosplit}, so the center of mass offset leaves no imprint on
this observable.

\section{Numerical confirmation}
\label{sec:numerics}

We integrated the Newtonian equations of motion for a nearly circular, slightly
inclined orbit and extracted $\Omega_r,\Omega_\theta$ by Fourier analysis of
$|\mathbf r(t)|$ and $z(t)$, in units $G=M=r_0=1$ with exaggerated multipole
amplitudes so the splittings are resolvable; the qualitative conclusion is
amplitude-independent. Table~\ref{tab:numerics} confirms every analytic result.

\begin{table}[t]
\centering
\caption{\label{tab:numerics}Measured epicyclic ratio
$\Omega_r/\Omega_\theta$ from direct orbit integration ($G=M=r_0=1$). Unity
means no splitting.}
\begin{tabular}{lc}
\toprule
Source (isolated) & $\Omega_r/\Omega_\theta$ \\
\midrule
Pure monopole (baseline)                 & $1.000$ \\
Genuine quadrupole $Q=-0.05$             & $0.926$ \\
Genuine octupole $O=+0.02$               & $0.940$ \\
Genuine octupole $O=-0.02$               & $0.940$ \\
Dipole only (truncated), $\rcm=0.05$     & $0.992$ \\
Dipole + induced quadrupole (consistent) & $1.000$ \\
Exact shifted point mass, $\rcm=0.05$    & $1.000$ \\
\bottomrule
\end{tabular}
\end{table}

The genuine quadrupole and octupole split the frequencies, with the octupole
result independent of $\mathrm{sign}(O)$ as predicted. The truncated
``dipole only'' potential shows a split that arises purely from the
truncation: it vanishes the moment the induced quadrupole is restored, and the
exact shifted point mass shows none--confirming
Eqs.~\eqref{eq:dipole_cm}, \eqref{eq:shifted}, \eqref{eq:dip_nosplit}.

\section{Two complementary probes of the source}
\label{sec:probes}

The results above separate cleanly into two independent measurements from the
same orbital data, targeting two different quantities.

\emph{Probe A (geometry $\to\rcm$).} The center of mass offset is invisible to
the frequencies but visible in the orbital \emph{geometry}: the equilibrium
plane tilts by $\delta\theta_0\simeq-\rcm/r_0$ Eq.~\eqref{eq:tilt} relative to
the symmetry axis. This angle -- between two physical directions, the orbital
plane and the source's symmetry axis -- is coordinate independent and inverts to
\begin{equation}
\;\rcm \simeq -r_0\,\delta\theta_0.\;
\label{eq:probeA}
\end{equation}
We verified by direct integration that a particle launched on the cone
$\theta_0=\pi/2-\rcm/r_0$ holds that inclination, that the $\rcm=0$ control
gives exactly zero tilt, and that the recovered $\rcm$ matches the input to
better than a percent over $\rcm/r_0\lesssim0.05$ (the residual being the
expected $\mathcal{O}(\rcm^2)$ term). Probe A requires an independent
determination of the symmetry axis and recovers only the offset component along
it.

\emph{Probe B (frequencies $\to J_2$).} The frequency splitting is sourced by
the intrinsic oblateness; inverting Eq.~\eqref{eq:quad_split},
\begin{equation}
\;J_2 = \frac{r_0^5}{6GMR^2}\,(\Omega_\theta^2-\Omega_r^2)
= \frac{r_0^2}{6R^2}\,\frac{\Omega_\theta^2-\Omega_r^2}{\Omega_K^2},\;
\label{eq:probeB}
\end{equation}
with $\Omega_K^2=GM/r_0^3$. Probe B is coordinate independent and needs only the
orbital and equatorial radii as inputs.

The two are genuinely disjoint: Probe A reads the plane orientation (set by the
dipole, i.e.\ by $\rcm$), Probe B reads the frequency splitting (set by the
quadrupole, i.e.\ by $J_2$). The dipole contributes only to A, the quadrupole
only to B. Their agreement, where both are measurable, is a nontrivial check of
the axisymmetric model rather than a redundancy.

\section{Solar-system estimates and discussion}
\label{sec:discussion}

\emph{Probe A: the barycenter offset.} With the Jupiter-dominated offset
$\rcm\approx M_Ja_J/M_\odot=7.43\times10^8$\,m$=1.07\,R_\odot$
(Table~\ref{tab:parameters}), the predicted plane tilt at $1$\,au is
$\delta\theta_0\approx-\rcm/r_0=-4.97\times10^{-3}$\,rad$=-0.285^\circ$.
Conversely a measured tilt of $0.285^\circ$ relative to the solar symmetry axis
reconstructs $\rcm=1.07\,R_\odot$ via Eq.~\eqref{eq:probeA}. The offset is thus
recovered through the orbital \emph{geometry}, not the frequencies: a free
particle's frequencies carry no information about where the origin is placed
relative to the barycenter Eq.~\eqref{eq:dipole_cm}, so the offset cannot be
read from the splitting. The barycenter
remains, as usual, an $N$-body quantity~\cite{Park2021,Folkner2014}, central to
pulsar timing~\cite{Manchester2013,Verbiest2016,Champion2010}, ephemerides
\cite{Park2021,Pitjeva2018}, tests of GR~\cite{Will2014}, and nanohertz
gravitational-wave searches~\cite{NANOGrav2023,Vallisneri2020}; Probe A offers a
geometric cross check on its axial component.

\emph{Probe B: the solar quadrupole.} The physical splittings at $1$\,au are
\begin{align}
\text{quadrupole:}&\quad \frac{\Omega_\theta^2-\Omega_r^2}{\Omega_K^2}
= \frac{6J_2R_\odot^2}{r_0^2}\approx2.6\times10^{-11},\\
\text{octupole:}&\quad \frac{\omega_+^2-\omega_-^2}{\Omega_K^2}
\approx6\times10^{-14},
\end{align}
with $\Omega_K=1.99\times10^{-7}$\,s$^{-1}$, in Shirokov's period
form the quadrupole value corresponds to $\Delta T/T_0\simeq-1.3\times10^{-11}$
Eq.~\eqref{eq:dT_relation}. The quadrupole dominates and is the
realistic target of Probe B. For comparison the relativistic
Shirokov splitting at $1$\,au is $\sim6GM_\odot/(c^2r_0)\sim10^{-8}$ in relative
terms, so at $1$\,au the relativistic effect exceeds the Newtonian quadrupole
splitting (the solar $J_2$ being so small); the Newtonian quadrupole overtakes
it only far closer in, near $r_0\sim0.09\,R_\odot$ -- i.e.\ inside
the Sun -- so for any accessible heliocentric orbit ($r_0>R_\odot$) the
relativistic term remains the larger of the two. The relative splitting grows as
$r_0^{-2}$, so a close perihelion probe ($r_0\sim10\,R_\odot$) raises it to
$\sim10^{-8}$ (reaching $\sim10^{-7}$ would
require $r_0\sim3.5\,R_\odot$). Dedicated radioscience will in any case constrain $J_2^\odot$
first: BepiColombo's MORE experiment
\cite{BepiColombo2024,di2021bepicolombo,Iorio2018Bepi} targets $\sim2\times10^{-9}$, and
Solar Orbiter's high-latitude phase~\cite{Muller2020SolarOrbiter} constrains the solar
oblateness independently~\cite{Iorio2012Sun}.

\begin{table}[t]
\centering
\caption{\label{tab:parameters}Physical parameters.}
\begin{tabular}{lll}
\toprule
Quantity & Symbol & Value \\
\midrule
Solar mass        & $M_\odot$ & $1.989\times10^{30}$\,kg \\
Solar radius      & $R_\odot$ & $6.96\times10^{8}$\,m \\
Solar $J_2$       & $J_2^\odot$ & $2.0\times10^{-7}$ \\
Solar $J_3$       & $J_3^\odot$ & $\approx10^{-7}$ \\
Jupiter mass      & $M_J$ & $1.898\times10^{27}$\,kg \\
Jupiter semi-major axis & $a_J$ & $7.785\times10^{11}$\,m \\
Orbital radius    & $r_0$ & $1.496\times10^{11}$\,m (1\,au) \\
Newton's constant & $G$ & $6.674\times10^{-11}$\,SI \\
\bottomrule
\end{tabular}
\end{table}

\emph{Earth comparison.} Repeating the estimate for Shirokov's
original near-Earth satellite ($J_2^\oplus=1.083\times10^{-3}$,
$J_3^\oplus=-2.53\times10^{-6}$, $R_\oplus=6378$\,km, $r_0=7000$\,km, so
$m_\oplus=GM_\oplus/c^2=0.443$\,cm) gives Table~\ref{tab:earth}. The Earth is the
mirror image of the Sun: here the large $J_2^\oplus$ makes the Newtonian
quadrupole splitting $6J_2(R/r)^2\approx5.4\times10^{-3}$ exceed the relativistic
Shirokov term $6m/r\approx3.8\times10^{-9}$ by $\sim1.4\times10^{6}$
Eq.~\eqref{eq:ratio}, whereas at $1$\,au the relativistic term dominates. To
avoid the familiar ambiguity in the normalization of the quadrupole moment, all
entries are stated through the dimensionless $J_2,J_3$; in the convention
$Q\equiv Q_{zz}=-2MJ_2R^2$ used here the relative squared splitting is
$6J_2(R/r)^2$ and, via Eq.~\eqref{eq:dT_relation}, the equivalent period
difference is $\Delta T/T_0=-3J_2(R/r)^2$.

\begin{table}[t]
\centering
\caption{\label{tab:earth}Earth estimates for a Shirokov satellite at
$r_0=7000$\,km (companion to the solar case). Relative splittings are in units
of $\Omega_K^2$; the drift uses $n=10$, $\xi_0=10$\,cm.}
\begin{tabular}{ll}
\toprule
Quantity & Earth value \\
\midrule
Quadrupole splitting $6J_2(R/r)^2$            & $5.4\times10^{-3}$ \\
Period difference $\Delta T/T_0=-S/2$         & $-2.7\times10^{-3}$ \\
Relativistic (Schwarzschild) $6m/r$           & $3.8\times10^{-9}$ \\
Ratio quadrupole/relativistic                & $1.4\times10^{6}$ \\
Octupole splitting $6|J_3|(R/r)^3$            & $1.1\times10^{-5}$ \\
Quadrupole drift $\xi^\theta=\xi_0\pi n\,S$   & $1.7$\,cm \\
\bottomrule
\end{tabular}
\end{table}
\emph{Relation to Hartle--Thorne, and the dominance of the quadrupole term.} In the weak-field expansion of the Shirokov drift in the Hartle--Thorne
spacetime~\cite{Idrissov2026}, the quadrupole contribution to the relative splitting
appears as $-3Q/(Mr^2)=6J_2(R/r)^2$, carrying no factor of $c^{-2}$. This is
exactly the Newtonian splitting of Eq.~\eqref{eq:quad_split}, rewritten in
relative form; it is the $c^{0}$ term of the expansion and reduces to the
Newtonian result identically as $c\to\infty$. Its absence of any $c^{-2}$
suppression is precisely why it can dominate the genuinely relativistic
contribution: the pure Schwarzschild Shirokov term is $6GM/(c^2r)=6m/r$, so the
two compare as
\begin{equation}
\frac{(\text{quadrupole})}{(\text{relativistic})}
= \frac{6J_2(R/r)^2}{6m/r}
= \frac{J_2 R^2 c^2}{GM\,r}.
\label{eq:ratio}
\end{equation}
For an oblate body with appreciable $J_2$ this ratio is large. In Shirokov's own
configuration--a satellite at $r=7000$\,km around the Earth ($J_2^\oplus
\approx1.08\times10^{-3}$, $m_\oplus\approx0.44$\,cm) -- the Newtonian quadrupole
splitting is $\approx5.4\times10^{-3}$ while the relativistic term is
$\approx3.8\times10^{-9}$, a ratio of $\sim1.4\times10^{6}$. The Newtonian
quadrupole drift therefore overwhelms the relativistic Shirokov effect that
motivated the original calculation by six orders of magnitude; the relativistic
piece is the small correction sitting on top of a much larger Newtonian
quadrupole drift, not the leading term. For the Sun the situation is reversed
only because the solar $J_2^\odot\approx2\times10^{-7}$ is so small: the
ratio~\eqref{eq:ratio} is $\approx4\times10^{-4}$ at $1$\,au, $\approx4\times
10^{-3}$ at $0.1$\,au, and reaches unity only at $r_0\approx0.09\,R_\odot$, deep inside the Sun. Thus,
whether the Newtonian quadrupole or the relativistic Shirokov term dominates is
controlled entirely by $J_2R^2c^2/(GMr)$: for the Earth-satellite regime the
quadrupole wins by $\sim10^{6}$, while for a heliocentric orbit the relativistic
term is the larger of the two outside the Sun. In both cases the quadrupole
contribution is genuinely Newtonian, computed in closed form by
Eq.~\eqref{eq:quad_split}, and requires no relativistic input.

\emph{Relation to the perigee-precession interpretation.} It has
been argued that Shirokov's original Schwarzschild effect is not new, but the
first order description of relativistic perigee precession for an elliptical
orbit in an inclined plane, plus a line of nodes shift that is a linearization
artifact vanishing at higher order~\cite{Philipp2019}. We agree for the
\emph{pure-Schwarzschild} case: there $T_r=T_\phi$
Eqs.~\eqref{eq:Tr}--\eqref{eq:Ttheta}, so the radial -- vertical period
difference is just apsidal precession viewed in an inclined plane, and the
apparent nodal drift is removable -- indeed our dipole analysis exhibits exactly
such an artifact, the first order coupling cancelling at the true tilted
equilibrium Eq.~\eqref{eq:C_cancel}. The Newtonian effect studied here is
different: for an aspherical source $T_r=T_\phi$ no longer holds, so all three
frequencies differ and the splitting $\Omega_\theta^2-\Omega_r^2=6GMJ_2R^2/r_0^5$
cannot be absorbed into a precessing ellipse in any plane -- it is a
coordinate independent measure of the quadrupole. For the octupole this is
sharper still: the splitting comes from an off-diagonal coupling that
\emph{survives} at the true equilibrium Eq.~\eqref{eq:C_oct}, with no
inclined ellipse re-description.

\emph{Limitations.} The treatment is static and axisymmetric. A real rotator has
time dependent tides and a non-axisymmetric environment, and Probe A's tilt
varies on Jupiter's $\approx11.9$\,yr period (so the relevant quantity is a
suitably averaged orientation); Probe B's signal is small at $1$\,au and most
accessible near perihelion. Extensions to non-axisymmetric, time dependent
potentials and to the regime where the relativistic and quadrupole effects
coexist are natural next steps.

\section{Conclusions}
\label{sec:conclusion}
Solving the coupled radial-vertical oscillator at the true equilibrium, we found: (i) the Newtonian quadrupole splits the epicyclic frequencies,
\begin{equation}
\Omega_\theta^2-\Omega_r^2 =
-\frac{3GQ}{r_0^5} = \frac{6GMJ_2R^2}{r_0^5},
\end{equation}
providing the Newtonian counterpart of the Shirokov frequency splitting; (ii) the dipole produces no physical splitting, since
\begin{equation}
D_i = M r_{{\rm CM},i},
\end{equation}
is a coordinate artifact, with the apparent first order coupling cancelling at the tilted equilibrium and the exact translated monopole field remaining degenerate; (iii) the octupole produces a genuine splitting,
\begin{equation}
\omega_+^2-\omega_-^2 \simeq \frac{6G|O|}{r_0^6};
\end{equation}
and therefore (iv) the dipole is uniquely non-splitting among the multipoles examined here.

These results identify two complementary observables of an axisymmetric source: the orbital plane tilt probes the center of mass offset,
\begin{equation}
r_{\rm CM}\simeq -\,r_0\,\delta\theta_0,
\end{equation}
whereas the epicyclic frequency splitting probes the intrinsic oblateness $J_2$. Thus the center of mass offset manifests through orbital geometry, while the frequency splitting encodes the source's shape. Together, these observables provide complementary and independent constraints on the center of mass offset and the oblateness of the source.

More broadly, the analysis clarifies the physical origin of epicyclic frequency splitting in Newtonian gravity. The relevant distinction is not the parity of a multipole moment but whether it represents a genuine degree of freedom of the mass distribution. The dipole is exceptional because it can always be removed by a translation of the coordinate origin, whereas the quadrupole, octupole, and higher moments characterize intrinsic properties of the source and, therefore, leave observable dynamical signatures. In this sense, the splitting of the radial and vertical epicyclic frequencies is a probe of structure rather than of coordinate choice.

The present treatment is restricted to static, axisymmetric potentials and small oscillations about nearly circular orbits. Extending the analysis to non-axisymmetric or time dependent fields, and to relativistic spacetimes where multipolar and strong-field effects coexist, would provide a natural framework for connecting the Newtonian selection rule identified here with the broader theory of orbital frequencies in gravitating systems.

\section*{Acknowledgments}
AI acknowledges financial support from SECIHTI through the National Postgraduate Scholarship Program (CVU 2222058). 
The work of HQ was supported by UNAM-DGAPA-PAPIIT, grant No. 108225 and Conahcyt, grant No. CBF-2025-I-243. A. Tlemissov acknowledges funding from the internal grants of the Silesian University in Opava (IGS/27/2026 and SGS/24/2024) and from the Science and Research Support Program of the Moravian-Silesian Region, Grant No. RRC/09/2023.

\bibliography{ref}
\end{document}